\def \SAIT #1 #2 {{\em Mem.\ Soc.\ Astron.\ It.\/} {\bf #1}, #2}
\def \MESS #1 #2 {{\em The Messenger\/} {\bf #1}, #2}
\def \ASTRNACH #1 #2 {{\em Astron. Nach.\/} {\bf #1}, #2}
\def \AAP #1 #2 {{\em Astron. Astrophys.\/} {\bf #1}, #2}
\def \AAL #1 #2 {{\em Astron. Astrophys. Lett.\/} {\bf #1}, L#2}
\def \AAR #1 #2 {{\em Astron. Astrophys. Rev.\/} {\bf #1}, #2}
\def \AAS #1 #2 {{\em Astron. Astrophys. Suppl. Ser.\/} {\bf #1}, #2}
\def \AJ #1 #2 {{\em Astron. J.\/} {\bf #1}, #2}
\def \ANNREV #1 #2 {{\em Ann. Rev. Astron. Astrophys.\/} {\bf #1}, #2}
\def \APJ #1 #2 {{\em Astrophys. J.\/} {\bf #1}, #2}
\def \APJL #1 #2 {{\em Astrophys. J. Lett.\/} {\bf #1}, L#2}
\def \APJS #1 #2 {{\em Astrophys. J. Suppl.\/} {\bf #1}, #2}
\def \APSS #1 #2 {{\em Astrophys. Space Sci.\/} {\bf #1}, #2}
\def \ASR #1 #2 {{\em Adv. Space Res.\/} {\bf #1}, #2}
\def \BAIC #1 #2 {{\em Bull. Astron. Inst. Czechosl.\/} {\bf #1}, #2}
\def \JSQRT #1 #2 {{\em J. Quant. Spectrosc. Radiat. Transfer\/} {\bf #1}, #2}
\def \MN #1 #2 {{\em Mon. Not. R. Astr. Soc.\/} {\bf #1}, #2}
\def \MEM #1 #2 {{\em Mem. R. Astr. Soc.\/} {\bf #1}, #2}
\def \PLR #1 #2 {{\em Phys. Lett. Rev.\/} {\bf #1}, #2}
\def \PASJ #1 #2 {{\em Publ. Astron. Soc. Japan\/} {\bf #1}, #2}
\def \PASP #1 #2 {{\em Publ. Astr. Soc. Pacific\/} {\bf #1}, #2}
\def \NAT #1 #2 {{\em Nature\/} {\bf #1}, #2}

\documentstyle[psfig]{memsait}
\begin{opening}
\title{BEPPOSAX OBSERVATIONS OF BRIGHT SEYFERT 2 GALAXIES:MEASURING THE 
INTRINSIC CONTINUUM EMISSION} 
\author{L. Bassani$^{1}$, M. Cappi$^{1}$, G. Malaguti$^{1}$,  
G.G.C. Palumbo$^{1,2}$,
M. Dadina$^{3}$, A. Comastri$^{4}$, G. Di Cocco$^{1}$, P.Blanco$^{5}$,
D. Dal Fiume$^{1}$, A. Fabian$^{6}$, F. Frontera$^{1,7}$,G.
Ghisellini$^{8}$, P. Grandi$^{9}$, M. Guainazzi$^{10}$, F. Haardt$^{11}$,
T. Maccacaro$^{8}$, R. Maiolino$^{12}$, G. Matt$^{13}$, L. Piro$^{9}$,
A. Santangelo$^{14}$, M. Trifoglio$^{1}$, N. Zhang$^{15}$}
\institute{
$^{1}$ Istituto Te.S.R.E./CNR, Bologna, Italy\\
$^{2}$ Dipartimento di Astronomia, Universita` di Bologna, Bologna, Italy\\
$^{3}$ BeppoSAX S.D.C., ASI, Roma, Italy\\
$^{4}$ Osservatorio di Bologna, Bologna, Italy\\
$^{5}$ UCSD, San Diego Ca, USA\\
$^{6}$ Institute of Astronomy, Cambridge University,
Cambridge, UK \\
$^{7}$ Dipartimento di Fisica, Universita' di Ferrara, Ferrara, Italy\\
$^{8}$ Osservatorio Astronomico di Brera, Milano, Italy\\
$^{9}$ Istituto Astrofisica Spaziale/CNR, Frascati, Italy \\
$^{10}$ ESTEC/SA, ESA, Noordwijk, The Netherlands\\
$^{11}$ Dipartimento di Fisica, Universita' di Milano, Milano, Italy \\
$^{12}$ Osservatorio Astrofisico di Arcetri, Firenze,  Italy\\
$^{13}$ Dipartimento di Fisica,Universita' di RomaIII, Roma, Italy\\
$^{14}$ IFCAI/CNR, Palermo, Italy\\
$^{15}$ University Space Research Association, Huntsville-Al, USA\\}
\date{} 
\end{opening}

\begin{document}

\oddpagefooter{}{}{} 
\evenpagefooter{}{}{} 
\ 
\bigskip
\normalsize

\begin{abstract}
We report broad band (0.1-200 keV) X-ray observations, made by BeppoSAX, of
a sample of bright Seyfert 2 galaxies: NGC7172, NGC2110, NGC4507, Mkn 3
and NGC7674.
These spectra provide  a better understanding of the effects of X-ray
reprocessing by cold material in the source and allow to put tighter 
constraints on the various spectral parameters involved. In particular, 
the data are used to determine, with  less
ambiguities than in the past, the shape of the intrinsic continuum emission
by means of the high energy data. Within the small sample both
Compton thin and Compton thick sources are found according to the 
expectations of the unified theory.

\end{abstract}

\section{Introduction}
The spectra of Seyfert (Sey) 2 galaxies detected in X-rays show absorption
by neutral matter significantly in excess of the galactic one along 
the line of sight (Smith and Done 1996, Turner et al. 1997a). On the contrary 
Sey1s do not exhibit evidence for intrinsic absorption by cold material.
This difference is interpreted as a further evidence in favour of the unification 
model, which states  that the discriminating parameter between the 
two Sey types is the observer's orientation with respect to an obscured 
molecular torus. It follows that the intrinsic spectra of Sey1s and 2s should
be the same once the effects of the torus absorbing material are
properly accounted for. We have tested this expectation against a sample of
5 Sey2s observed by BeppoSAX during the AO1: NGC2110, NGC7172, NGC4507, 
Mkn 3 and NGC7674. In this contribution preliminary results are presented with 
emphasis on the role of the various spectral components.

\section{Seyfert 2 broad band spectral characteristics}
The 2-10 keV band is the one  most effected by the torus material both in terms
of absorption and reprocessing. In sources 
having N$_{\rm H}$  $<$10$^{24}$ cm$^{-2}$, the
matter is optically thin and the intrinsic emission should be visible 
above a given energy (generally a few keV); 
when N$_{\rm H}$ is around a few 10$^{24}$ cm$^{-2}$ 
only X-rays in the 10-100 keV  region can pass through the torus. Beside
absorbing the primary emission, the torus is also responsible for reflecting 
the continuum (Ghisellini et al. 1994); thus a 
bump in the 10-50 keV region similar to what observed in Sey1 is expected.
An iron line produced either
via transmission through (Leahy and Creighton 1993) and/or scattering-
reflection (Ghisellini et al. 1994) by the absorbing material should also be
observable: expected 
values of the iron line equivalent widths (EW) range from 
100 eV up to $\sim$ 1keV for high enough columns, i.e N$_{\rm H}$
$\ge$ 10$^{23}$ cm$^{-2}$. 
For  column densities in excess of  10$^{25}$ cm$^{-2}$, the absorbing matter
is Compton thick and the intrinsic continuum is not directly visible; in this
case only the (unabsorbed) emission reflected by the torus (but also scattered in 
warm material above the nucleus) should be  detected by the observer and 
the iron line being measured against a depressed continuum is expected to have
an EW  $\ge$ 1keV (Maiolino et al. 1998). Moreover, a (absorbed) reflection 
component and associated iron fluorescence line due to an accretion
disk (as observed in Sey1s , see Perola et al. this proceeding) should be 
present in all Sey2s. However, at this stage of the analysis,  
this component has been
neglected for the sake of simplicity and also in view of the fact
that in Sey2s it should be a second order effect because of the edge-on geometry and 
of the high N$_{\rm H}$ involved.
Table 1 lists for the observed sample of Sey2s the spectral
parameters  deduced by fitting their broad band X-ray spectra with a baseline
model which includes an absorbed power law, a reflection component due the 
torus and a narrow Fe K gaussian line; to account 
for the low energy data  another power law having the photon index thight 
to the first one is also included in the fit. 
The uncertainties quoted 
correspond to 90$\%$ confidence for two interesting parameters.
One striking result of Table 1 is the wide range of values observed for the 
parameter R , i.e the relative amount of reflection compared to the directly 
viewed primary power law. We range from sources having  no or 
little reflection like NGC2110 and NGC4507  to objects reflection  dominated 
like Mkn 3 or even more NGC7674. In this latter source the primary emission 
is totally absorbed and only radiation reprocessed by the torus is seen
making the object a likely Compton thick candidate 
(see Malaguti et al. 1998 for details).
While broadly speaking the wide range in R is expected in the unified model,
as both Compton thin and thick sources are predicted, the  
observation of low reflection in some sources requires either that the
torus is  intrinsically optically thin or that it covers a small
angle at the source.
In this case, the small reflection observed maybe compatible with an origin in the 
accretion disk viewed edge-on; at least for the case of NGC2110 this 
interpretation would be in agreement with the observation of a broad relativistic 
component in the iron line (Weaver $\&$ Reynolds 1998). 
As expected, an iron line at 6.4 keV 
is clearly detected in each source and consistent with theoretical
expectations.
In Compton thin cases, the line is compatible 
with transmission in the torus (NGC2110 and NGC4507) possibly coupled with 
a reflection line component (NGC7172). In Mkn 3, the absorption
is high enough to depress the continuum significantly leading to a line
with EW $\sim$ 1 keV which is the result of both transmission and reflection 
in the torus (Cappi et al. 1998).
In NGC7674, as the source is Compton thick, 
the line is simply the result of reflection in the torus (Malaguti et al. 1998). 

\begin{center}
{\bf Table 1: Seyfert 2 Galaxies in the sample}\\
\begin{tabular}{ l  c  c c c c c }
\hline
Source Name& N$_{\rm H}$$^{*}$ & EW$_{\rm K\alpha}$ $^{\ddagger}$   &R&
$\alpha$&
F$_{\rm 2-10keV}$$^{\dagger}$ & F$_{\rm 20-100keV}$$^{\dagger}$  \\
\hline
NGC2110 &    4.1${\pm 0.2}$  & 180${\pm30}$ & $<$0.2  &  1.7${\pm0.1}$   &3.0 &5.8 \\

NGC7172$^{N}$     &11.5$^{+12.5}_{-10.7}$      &224$^{+650}_{-54}$
&1.0 $^{+2.7}_{-0.4}$      &1.9${\pm0.3}$    &1.0 & 3.3 \\

NGC4507$^{N}$     &70$^{+15}_{-15}$           &120$^{+192}_{-68}$
&0.40 $^{+0.2}_{-0.2}$  &2.0${\pm0.1}$    &3.6 &15.0 \\

Mkn 3        &130$^{+15}_{-25}$        &997$^{+300}_{-307}$
&0.9 $^{+0.1}_{-0.1}$      & 1.8${\pm0.1}$   &0.65 &11.3  \\

NGC7674     &$>$1000                 &900$^{+470}_{-299}$
&$>$ 40        &1.9${\pm 0.2}$     &0.05 & 1.0 \\
 \hline
\end{tabular}
\\
\end{center}
$^{*}$ in units of 10$^{22}$ cm$^{-2}$,
$^{\ddagger}$ in units of eV,
$^{\dagger}$ observed flux in units of 10$^{-11}$ erg cm $^{-2}$ s$^{-1}$\\
$^N$  preliminary results to take with care.
\\

In all our sources, the values of the intrinsic power law photon 
index are similar and well constrained in the range 1.7-2.0; therefore
the primary component of the emission bears no differences from  
what observed in Sey1s (see Perola et al. this proceedings).
This is also true for those sources previously reported as flat, i.e
NGC2110 and NGC7172 (Smith and Done 1996 and Guainazzi et al. 1998)
highlighting the crucial role of the PDS in estimating the primary 
power law component. To put a more stringent constrain on the
intrinsic slope, a mean Sey2 spectrum has been obtained by
summing all data in the 20-100 keV range. Note that the absorption affecting
Mkn 3 is negligible above 20 keV thus allowing this source to be added too;
viceversa NGC7674 has not been considered for its different spectral shape
in the PDS range.The resulting spectrum is shown in figure 1; a simple power
law fit to the data gives a photon index of 1.81 $\pm$ 0.05.
We have also tested individual source spectra for a high energy cut-off. 
In the case of Sey2s this is not straightforward due to the complexity
involved in the spectral fitting. Nevertheless, our data  suggest that
also in type 2 objects the typical cut-off energy is above 100 keV
as observed in Sey1s (Perola et al. this proceedings); for example in Mkn 3 
the lower limit on the cutoff energy is 150 keV at $\gg$ 99\% 
confidence (Cappi et al. 1998).

\section{Conclusions}
Preliminary BeppoSAX results on a small sample of bright Sey2 galaxies,
while confirming the basic expectations of the unified theory,
i.e. a primary shape similar to Sey1s but distorted by absorption
and reprocessing in the torus cold material, also highlights
the need for a better understanding of the role of the torus  
in these sources. Hopefully this will be possible by
enlarging the sample with data from more sources covering a broader
range of the space parameters.


\begin{figure}
\hspace{13.5cm}
\psfig{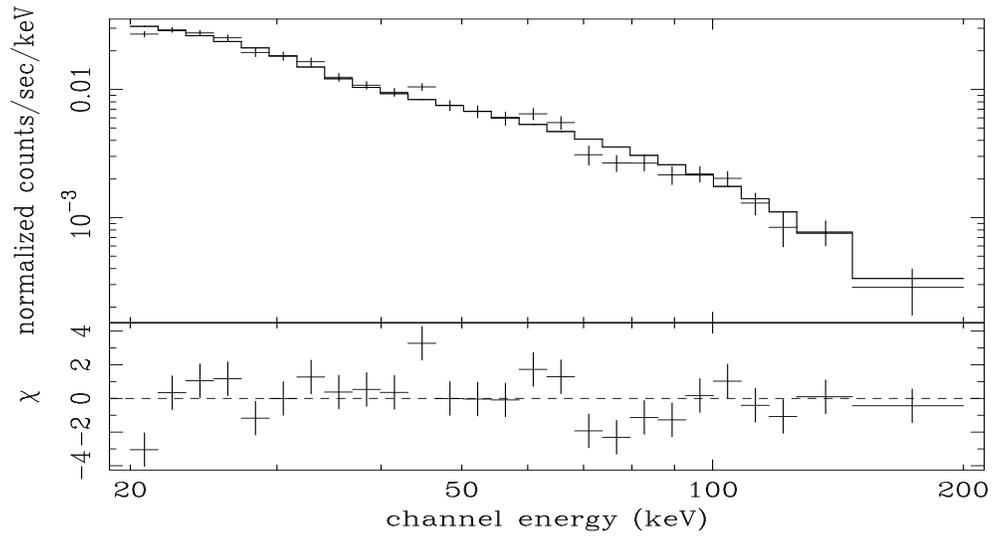}
\caption[h]{Mean Sey2 PDS spectrum (20-200 keV).}
\end{figure}





\end{document}